\documentclass[preprint,showpacs,showkeys]{revtex4-1}
\usepackage{graphicx}
\makeatletter
\newcommand{\Rmnum}[1]{\expandafter\@slowromancap\romannumeral #1@}
\makeatother
\newcommand{\dbar}{d\mkern-5mu\mathchar'26}
\begin{document}
 \title{A Note on Non-equilibrium Work Fluctuations and  Equilibrium Free Energies}
\author{M. Suman Kalyan${}^{\dagger}$, G. Anjan Prasad${}^{\star}$, V. S .S. Sastry${}^{\dagger}$
and K. P. N. Murthy${}^{\dagger}$}
\affiliation{${}^{\dagger}$ School of Physics,
University of Hyderabad,
Central University P.O.,
Hyderabad 500 046.
Andhra Pradesh, India\\
${}^{\star}$ Institute for Computational Physics,
Stuttgart University, Pfaffenwaldring 27,
70569 Stuttgart
Germany}
\date{\today}
\begin{abstract}
We consider in this paper,  a few important issues in
non-equilibrium work fluctuations and their relations to
equilibrium free energies. First  we  show that  Jarzynski
identity can be viewed as a cumulant expansion of  work.
For a switching process which is nearly quasistatic the
work distribution is  sharply peaked and Gaussian.
We show analytically that  dissipation given
by average work minus  reversible work  $W_R$,
decreases when the process becomes more and more quasistatic. Eventually, in 
the quasistatic 
reversible limit,  the dissipation 
vanishes. However
estimate of $p$ - the probability of violation of the
second law
given by the integral of the tail of the work
distribution from $-\infty$ to $W_R$, increases and takes a value of
$0.5$ in the quasistatic limit.
We show this analytically employing Gaussian integrals
given by  error functions and
Callen-Welton theorem that relates fluctuations to dissipation in
process that is nearly quasistatic.
 Then we carry out  Monte Carlo
simulation of non-equilibrium processes
in a liquid crystal system in the presence of an electric field and
present results on reversible work, dissipation, probability of violation of
the second law and distribution of work
\end{abstract}

\maketitle

\section{Introduction}
Non equilibrium work fluctuations and equilibrium free energy differences
are related to each other through a remarkable identity proposed and
proved  by Jarzynski \cite{CJ_1,CJ_2}. Ever since several studies have been carried
out to establish the validity and utility of the identity for various
non-equilibrium evolutions - stochastic, see {\it e.g.} \cite{J_Kurchan,RJHGMS}
as well as  deterministic, see {\it e.g.} \cite{CJ_1,D_J_Evans}

We show in this paper that Jarzynski identity can be viewed as a cumulant
expansion. When we retain only the first cumulant in the expansion,  we get
the well known thermodynamic identity equating the work done on the system
to free energy change in a quasistatic reversible process.
With only the first two cumulants in the expansion, we get
Callen-Welton theorem \cite{HBCTAW}. When  we  include the
third and higher order cumulants we get  Jarzynski  identity.
We derive an analytical expression for
probability of
violation of the second law,
under Gaussian approximation to work fluctuations.
We then carry out Monte Carlo simulation of a
liquid crystal system in the presence of a varying electric field
and report numerical results on free energy changes,
dissipation, probability of violation of the second law and distribution
of work.

The paper  is organized as follows. We
start with a  brief introduction to a few relevant basic issues
of   heat and work in a quasistatic reversible process.
Then we consider heat and work  in  the context of irreversible processes.
This is followed by a discussion on the relation between
non-equilibrium work fluctuations and equilibrium free energy differences.
We show analytically that   the  probability of violation of the second law
increases with increase of time duration of the switching experiment
and in the asymptotic limit of a quasistatic process
it goes to one-half.
However dissipation
defined as average work minus reversible work
becomes smaller when the  duration of time of the switching process increases
and in the quasistatic limit it goes to zero.
We demonstrate the usefulness of Jarzynski identity \cite{CJ_1,CJ_2}
on a system of liquid crystals in the presence of external electric field.
We consider a lattice model  with each lattice site holding
a headless spin.
We calculate the work done when the external electric field is switched
from an initial value to a final
value following a well specified experimental protocol. We carry Monte Carlo
simulation of the non-equilibrium switching process
and collect an ensemble of work values. From this ensemble
we extract equilibrium properties of the liquid crystalline
system employing both Jarzynski identity and
Callen-Welton theorem \cite{HBCTAW}.
Principal conclusions of the study are
brought out briefly in the concluding section.
\section{Heat and Work : Reversible Processes}
Consider a  closed system in equilibrium at temperature $T$.
It draws a  small quantity $\dbar Q$  of heat by a
quasistatic reversible
process at constant temperature.
$\dbar Q$ is not a perfect differential.
However,
\begin{eqnarray}\label{Clausius_entropy}
dS =\frac{\dbar Q }{T},
\end{eqnarray}
is a perfect differential
where $S$ is entropy. Thus, during a quasistatic
reversible process,   entropy of the  system
increases
by $\dbar Q/T$.

To calculate  the work done on the system,
we start with the first law of thermodynamics, stated as,
\begin{eqnarray}\label{First_Law}
\dbar W=dU- \dbar Q,
\end{eqnarray}
where $\dbar W$ denotes the  work done on the system and $U$, the internal energy.
$\dbar W$ is not a perfect differential.

The first law is about conservation of energy and is valid for all processes,
quasistatic, non-quasistatic, reversible,
 irreversible or  otherwise. However, if the process is quasistatic and reversible,
  then we can replace
$\dbar Q$ by $TdS$, see Eq. (\ref{Clausius_entropy}),  and write,
\begin{eqnarray}\label{First_Law_T}
\dbar W=dU-TdS.
\end{eqnarray}
If the process is also
isothermal, we have
\begin{eqnarray}
\dbar W=d(U-TS).
\end{eqnarray}
We identify the term $U-TS$  as Legendre transform of the fundamental equation
$U\equiv U(S,V)$ where we transform the variable $S$ in favor of the \lq slope\rq\
\begin{eqnarray}
T(S,V) & = & \left( \frac{\partial U}{\partial S}\right)_{V}
\end{eqnarray}
and the variable $U$ in favor of the \lq intercept\rq\
\begin{eqnarray}
F(T,V) & = & U(S,V) - TS.
\end{eqnarray}
$F(T,V)$  is called Helmholtz free energy or
simply free energy.
Thus the work done equals  free energy change :
\begin{eqnarray}\label{W_dF}
 \dbar W=dF.
\end{eqnarray}
 For purpose of distinguishing process variables (heat and work)  
from state variables (energy $U$, 
entropy $S$ and free energy $F$)
we have considered in the above an infinitesimal process. $\dbar Q$ and $\dbar W$ 
are not perfect differentials; $dU$, $dS$ and $dF$ are perfect differentials. 
However the 
relation between free energy and work, see Eq. (\ref{W_dF}),  holds good for a 
quasistatic reversible process that  takes the system from one equilibrium state,
say $A$ 
to another equilibrium state, say $B$. We have
\begin{eqnarray}
 W = F(B) - F(A)
\end{eqnarray}
We call the above as reversible work and denote it by $W_R$
\section{Heat and Work :  Irreversible Processes}
To obtain a relation between change in free energy and work done when
the process is not reversible,
we start with the second law  inequality,
see
\cite{JSD} for an elegant proof,  given by
\begin{eqnarray}
dS\ >\ \frac{\dbar Q}{T}.
\end{eqnarray}
The above  implies
\begin{eqnarray}
\dbar Q\ <\ T\ dS
\end{eqnarray}
for an irreversible process.
Substitute this in Eq. (\ref{First_Law}) and get,
\begin{eqnarray}
\dbar W & > & dU - T dS
\end{eqnarray}
which for an isothermal irreversible process
reduces to
\begin{eqnarray}
\dbar W\ >\ dF.
\end{eqnarray}
Consider a process that takes the system irreversibly from an equilibrium state $A$ to 
another equilibrium state $B$. Let $W$ denote the work done during the process and 
$\Delta F = F(B) - F(A)$. 
We have
\begin{eqnarray}
W & \ >\  & \Delta F
\end{eqnarray}
 The work done exceeds  free energy change, if the
process is not reversible.
This is the best conclusion we can draw  from purely thermodynamics
considerations. 

A natural question that arises in this context
is about the meaning of the statement $W > \Delta F$.
A given  process can be realized
either in an experiment or  in a computer simulation.
We recognize that  work done, in general, would  differ from one
experiment or computer realization
of the  process to another.
Let us consider a process in which  a parameter
of the thermodynamics system
is switched from one value to another,  as per
a well defined experimental protocol. For example,  we
can change the volume of a system from one value
to another uniformly over a fixed duration of time $\tau$.
We call this a switching process. In general,  different
switching experiments, all carried out with the
same protocol,  will give rise to different values of $W$.
Only when the switching is done quasistatically and
reversibly does one get the same value of $W$ in
all experiments. Hence, in general,  we have
to deal with an ensemble
of values of $W$ and not just
a single value. Let us denote this ensemble
by $\Omega = \{ W_i\}$. It is then quite possible that  there
can exist realizations for which $W$ is
less than $dF$, thereby  {\it violating}
 the second law.  This should not surprise us since
even in the early days of statistical
mechanics,  Maxwell \cite{Maxwell}
correctly  recognized
that the second law is of statistical origin and hence there is
a non-zero probability of it being contravened.
The demon he  created
to drive home this  point, haunts us even today,
see {\it e.g.} \cite{Maxwell_Demon}.

Let the ensemble $\Omega$ of work values 
 be formally described by the distribution
$\rho(W;\tau)$ where $\tau$ denotes the duration of time over
which the process takes place; $\tau$ is  the switching time.
Typically we switch a macroscopic parameter, denoted by the symbol
$\Lambda$ from  an initial value say $\Lambda_I$  
to a final value say $\Lambda_F$.
This switching  can be carried out in any way.
The discussions below and the Jarzynski identity 
described  in the next section
hold good for any protocol of switching. 
However 
we consider the switching to  take place at constant rate, 
over a pre specified duration of time,  $\tau$.
Then by considering different values of $\tau $ we can describe 
different switching scenarios: 
if $\tau$ is small, we have fast switching; if $\tau$ is large we have 
slow switching; in the limit of 
$\tau$ $\rightarrow$ $\infty$, we have  a quasistatic reversible switching. Thus
\begin{eqnarray}
\Lambda(t)=\Lambda_I + (\Lambda_F - \Lambda_I)\times \frac{t}{\tau}, 
\ \ \ \ 0\ \le\ t\ \le\ \tau .
\end{eqnarray}
The probability of violation of the second law, denoted by the symbol $p(\tau)$
 is formally
given by the following integral.
\begin{eqnarray}\label{p_tau_formal}
p(\tau) &= &\int_{-\infty}^{W_R}dW\ \rho(W;\tau),
\end{eqnarray}

We note that for every thermodynamic variable there
corresponds a random variable in statistical mechanics.
The average of the random variable
over a suitable ensemble gives the value of the thermodynamic
variable. For example
$\langle E\rangle$   equals the
thermodynamic energy $U$, where the angular brackets  denote
average  over a canonical ensemble.   Hence, strictly   we should
state the second law as
\begin{eqnarray}
\langle\ W\ \rangle \ \geq\ dF\ .
\end{eqnarray}
In the above, the angular brackets denote
an average over an ensemble of switching experiments,
all carried out with the same protocol. In other words it is given by
\begin{eqnarray}
\langle W \rangle & = & \int_{-\infty}^{\infty} dW\ W\ \rho(W;\tau)
\end{eqnarray}
Stated thus,
the second law can never be violated.
We   define
\begin{eqnarray}\label{dissipation}
W_d  &=& \langle W\rangle - W_R
\end{eqnarray}
 as  dissipation and state the second law as
described  in the next section
\begin{eqnarray}
 W_d \ \ge\ 0
\end{eqnarray}
 for all processes. In the above, equality obtains
when the process is quasistatic and reversible.

If $\tau$ is very large but  not infinity,
$\rho(W;\tau)$ would be  sharply peaked and  Gaussian.
Consider a process that takes a system from an equilibrium state
$A$ to another equilibrium state $B$. Let the process be nearly
quasistatic. We have $\Delta F= F(B)-F(A)=W_R$. 
The time taken for the process, $\tau$,  is large.
For such a process, dissipation is
proportional to fluctuations,
and from Callen-Welton theorem
\cite{HBCTAW}
 we get,
\begin{eqnarray}\label{W_R_FDT}
\Delta F  = W_R  = \zeta_1 - \frac{1}{2}\  \beta\ \zeta_2
\end{eqnarray}
where
\begin{eqnarray}
\zeta_1  =  \langle W\rangle
\end{eqnarray}
  is  the  first  cumulant
and
\begin{eqnarray}
\zeta_2  =  \sigma^2_W=\langle W^2\rangle - \langle W \rangle ^2
\end{eqnarray}
is the second cumulant (or
variance) of
$W$. Thus we have $ W_d   = \beta\zeta_2/2$ from Eq. (\ref{W_R_FDT}):
dissipation ($W_d$) is proportional to fluctuation ($\zeta_2$ or $\sigma^2_W$).

Measuring energy in units of $k_BT=\beta^{-1}$,  we get,
from Eq. (\ref{W_R_FDT}),
\begin{eqnarray}\label{two_cumulants}
 -\beta\  dF =  (-\beta)\ \zeta_1 + \ \frac{1}{2!}\ (-\beta)^2\ \zeta_2\ .
\end{eqnarray}
The above relation is for a process which is nearly quasistatic. The distribution of work
for such a switching protocol is Gaussian. 
For a Gaussian, the third and higher order cumulants are all identically zero. 
We contend  that in a general process, which is not necessarily quasistatic or near quasistatic,
there should be additional terms in the Right Hand Side (RHS) of the above equation involving third 
and higher cumulants. Once we include these higher cumulants we get 
 Jarzynski identity, as shown in the next section.
\section{Non-equilibrium Work and Equilibrium Free Energies}
Including the terms involving the third and higher order
cumulants in the  RHS of Eq. (\ref{two_cumulants}), we get,
\begin{eqnarray}
 -\beta\  dF =(-\beta)\ \zeta_1   +  \frac{(-\beta)^2}{2!}\ \zeta_2
 + \sum_{n=3}^{\infty}\frac{(-\beta)^n}{n!}\ \zeta_n .
    \end{eqnarray}
We recognize immediately the Right Hand Side (RHS)
of the above equation as the cumulant expansion
of $W$.

 Let $\chi(\beta)$ denote the cumulant generating function. Thus we have,
\begin{eqnarray}
\chi(\beta) &=& -\beta dF
\end{eqnarray}
 or equivalently
\begin{eqnarray}
\exp (\chi)  &= & \exp(-\beta dF).
\end{eqnarray}
The moment generating function of $W$ is defined as
\begin{eqnarray}
\phi(\beta) & = & \int_{-\infty}^{+\infty}dW\ \exp(-\beta W)\ \rho(W;\tau)\nonumber
\\
   & & \nonumber\\
            & = & \langle\ \exp(-\beta \ W)\ \rangle .
\end{eqnarray}
The moment and cumulant generating functions are
related to each other. We have
\begin{eqnarray}
\chi(\beta)=\log \left[ \phi(\beta)\right].
\end{eqnarray}
We  see immediately that Jarzynski identity
given by,
\begin{eqnarray}
\langle\  \exp(-\beta W)\ \rangle = \exp(-\beta dF),
\end{eqnarray}
follows naturally from this.
\section{Gaussian  Work Distribution}
 We shall derive  an expression for the probability of
violation of the second law formally
given by Eq. (\ref{p_tau_formal}).
 For large $\tau$  the work distribution $\rho(W;\tau)$
would  be a sharply peaked
Gaussian with mean
$\langle W(\tau)\rangle$ and variance
$\sigma^2_W(\tau)=\langle W^2(\tau)\rangle -\langle W(\tau)\rangle ^2$.
Substituting the Gaussian in Eq. (\ref{p_tau_formal}) for $\rho(W;\tau)$ we get,
\begin{eqnarray}\label{poftau}
 p(\tau)
            &= & \frac{1}{\sigma_W(\tau)\sqrt{2\pi}}\int_{-\infty}^{W_R}dW
                 \exp\left( -\frac{1}{2} \frac{\left[ W-\langle W(\tau)\rangle\right]^2}
{\sigma^2_W(\tau)}\right)\nonumber\\
 & & \nonumber\\
& & \nonumber\\
&= & \frac{1}{\sqrt{\pi}}\int_{-\infty}^{{\displaystyle \frac{W_R-\langle W(\tau)\rangle }{\sigma_W(\tau)\sqrt{2}}}} d\xi\exp(-\xi^2)\nonumber\\[2mm]
  & & \nonumber\\
& & \nonumber\\
  & = &\frac{1}{2}-\frac{1}{2}\ {\rm erf}\left( \frac{\langle W (\tau)\rangle -W_R}{\sigma_W(\tau)\sqrt{2}}\right),
\end{eqnarray}
where,  the error function is defined \cite{errf} as,
\begin{eqnarray}
{\rm erf} (x) &=& \frac{2}{\sqrt{\pi}}\int_0^x d\xi\exp(-\xi^2)
\end{eqnarray}
For large $\tau$,  we have Callen-Welton theorem \cite{HBCTAW}
which tells us
\begin{eqnarray}
\langle W(\tau)\rangle  -W_R = \frac{1}{2}\beta\sigma^2_W(\tau).
\end{eqnarray}
Substituting  for dissipation in terms of fluctuations in Eq. (\ref{poftau}) we get
\begin{eqnarray}\label{p_CW}
p(\tau) & = & \frac{1}{2} - \frac{1}{2}{\rm erf}\left( \frac{\beta\sigma_W(\tau)}{2\sqrt{2}}\right).
\end{eqnarray}
Note ${\rm erf}(\infty)=1$ and ${\rm erf} (0)=0$.
From Eq. (\ref{p_CW}) we see that  $p(\tau)$ increases when $\tau$ increases.
In the limit of $\tau\to\infty$, $p(\tau)$ equals one-half.

This result can be understood as follows\cite {AR}. In the quasistatic limit of $\tau\rightarrow\infty$,  
$\rho(W; \tau)$
 becomes more and more sharply peaked 
such that $\langle W\rangle\rightarrow W_R$ and $\sigma_W\rightarrow 0$. 
By Callen-Welton
theorem $W_d \ \propto\ \sigma^2_W$. This implies that 
$\sigma_W\ \gg\ W_d$ 
as $W_d\ \rightarrow\  0$. Thus $W_d$ and 
the $\sigma_W$ both tend to zero, with $\sigma_W\ \gg\ W_d$. In other words
$W_d\rightarrow 0$  faster than 
$\sigma_W$ $\rightarrow 0$. This leads to $p(\tau)\ \rightarrow\ \frac{1}{2}$ 
 in the reversible limit.

However,
dissipation is defined  as
$W_d=\langle W\rangle -W_R$. In the reversible limit of $\tau\to\infty$, 
we have  $\langle W\rangle = W_R$ and hence  dissipation is  zero.

\section{Work fluctuations in liquid crystalline system}
Let us  now illustrate the above
on a lattice model of a liquid crystalline system.
To this end, we  consider an $L\times L\times L$
cubic lattice. Each lattice site holds a headless spin.
The spins on  nearest neighbor
lattice sites interact with each other {\it via}
Lebwohl-Lasher potential \cite{LL} see below. Besides, each spin
interacts with the external electric field, which is taken as the switching
parameter. Without loss of generality
we take the external field,
to be in the  $z$ axis direction and switch its magnitude.
The Hamiltonian for the nematic system is given by,
\begin{eqnarray}
H =-J\sum_{\langle i,j\rangle} P_2(\cos \theta_{i,j})-\frac{E^2}{2}\sum_i P_2(\cos \theta_i).
\end{eqnarray}
In the above, the symbol $\langle i,j\rangle$ denotes that
$i$ and $j$ are nearest neighbor lattice sites. The sum
runs over all
distinct pairs of nearest neighbor sites. We have employed periodic 
 boundary conditions along $x$, $y$ and $z$ coordinates. $\theta_{i,j}$
denotes the angle between
the spins and $J$ measures the strength of interaction.
In the simulation we express energy in units of $J$ and hence set $J=1$.
$E$ is the amplitude of the
external electric field and $\theta_i$, the  angle between
the spin at lattice site $i$ and the
external electric field.  $P_2(\eta)$ is the second Legendre
polynomial given by
\begin{eqnarray}
P_2(\eta) & = & \frac{3\cos^2 (\eta)-1}{2}
\end{eqnarray}
We carry out Monte Carlo simulation of the
response of the system to a process of switching $E$ from
say $E_0$  to a value of $E_f$.

\section{Monte Carlo simulation of  the switching process}
Start with an arbitrary initial microstate (spin configuration)  and employing
Metropolis algorithm \cite{Metropolis},  equilibrate the system at the desired value of $\beta$ with $E=E_0$.
Select randomly a microstate from the equilibrium ensemble. Calculate the energy.
Keep the micro state the same and switch the field from
$E_0$ to $E_1=E_0+\Delta E$.  This is called the work step.
The resulting change in energy,  called  work is
denoted by $\omega_1$.
Then implement a heat step {\it via} Metropolis algorithm
over one Monte Carlo sweep as described below.

Select randomly a spin and change its
orientation by a random amount. This can done by the procedure suggested by Barker and Watts \cite{Barker}.
Call the resultant microstate as trial state. If the energy of the trial state is less than that of the
current state accept the trial state as the next microstate in
the Markov chain. If the trial state energy is more, then
calculate $p=\exp[-\beta\Delta \epsilon]$ where $\Delta \epsilon$ is the
energy of the trial state minus that of the current state.
Draw a random number (uniformly and independently
distributed in the range zero to unity) and if it less than $p$
accept the trial state; otherwise reject the trial state and take
 the current state as the next state in the Markov Chain.
Carry out the above rejection/acceptance step $L^3$
number of times and this constitutes a Monte Carlo sweep.
The heat step is followed by a work step on the micro state
obtained at the end of the heat step.
During the work step the
field changes from $E_1$ to $E_2=E_1+\Delta E=E_0+2\Delta E$.
The work done is $\omega_2.$
The work and heat steps are repeated until
$E$ attains a
pre - determined value, say $E_f$.

Let $n$  be the number of work steps required to
switch the field from an initial value of $E_0$ to a final value $E_f$.
In other words $E_n=E_f=E_0+n\ \Delta E$, where
$\Delta E = (E_f-E_0)/n$.
The switching time $\tau$ is thus given by $n$.
We have
\begin{eqnarray}
W=\sum_{i=1}^n \omega_i\ .
\end{eqnarray}
We carry out the simulation independently
 for a large number of times with the same switching protocol
and construct a work ensemble $\{ W_i\}$ from which
all the required statistics
described below are calculated.
 In the simulation we have taken $E_0=0$ and $E_f=2$.
 The results presented in this paper are for  $L=3$.
 Jarzynski identity is valid even for systems which are small.
 For a small system there would be large fluctuations and 
 hence illustrating various issues, like relating work fluctuations to 
 free energy differences and to dissipation becomes an easy task. 
 This is the reason we have kept the system size small in this work.
 The size of the Monte Carlo ensemble of work values generated is one million. 

\section{Results and discussions}
Fig. (1) depicts results on free energy difference (or reversible work)
calculated from Jarzynski identity as a function of $\tau$
with $\beta^{-1} = 1.5$.

\begin{figure}[htp]
\includegraphics[width=140mm]{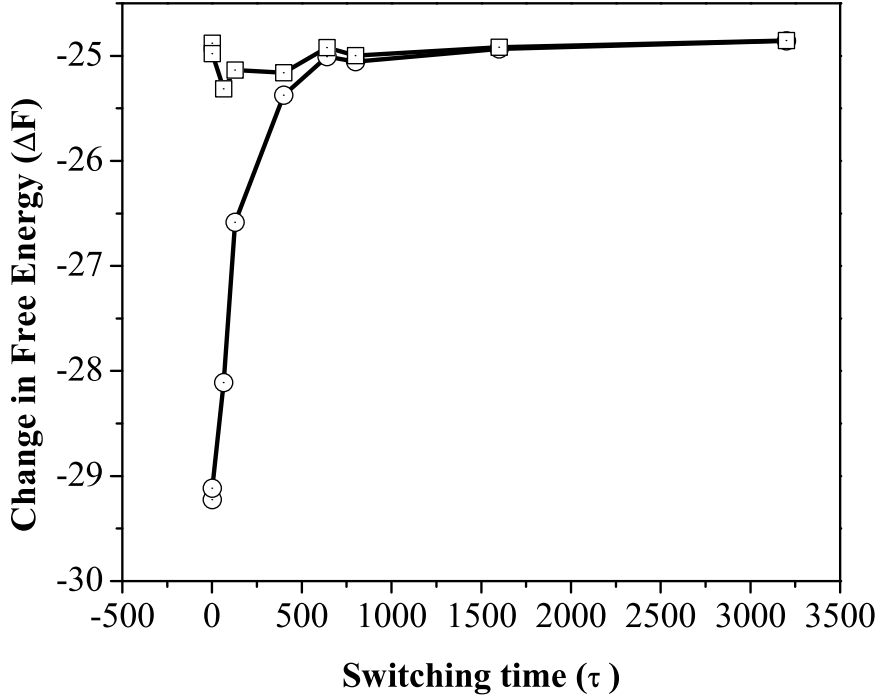}
\caption{\small{\it Free energy difference $\Delta F$
{\it versus} switching time $\tau$. The upper curve (squares)
denotes $\Delta F^J$  calculated from Jarzynski identity. The lower curve
(circle) denotes $\Delta F^{CW}$, calculated
 from Callen-Welton theorem.}}
\end{figure}

\noindent
We have also plotted in the
same graph the free energy change from Callen-Welton theorem.
As expected Callen-Welton theorem does not predict
free energy change correctly for small switching time
when the system is driven far from equilibrium. As $\tau$ increases the results from
Callen-Welton theorem converges to that from Jarzynski identity.

In Fig. (2) we
have plotted  dissipation given by  Eq. (\ref{dissipation})
and denoted as $W^J_d$,
 as a function $\tau$.
This has been calculated from the ensemble of work values
generated by the Monte Carlo simulation of the switching process.
The dissipation is large for small $\tau$. This is because
the system is driven far from equilibrium.
However as the switching time increases the process
becomes more and more quasistatic.
Dissipation decreases. In the limit of $\tau\to \infty$ it goes to zero.
For large $\tau$ dissipation can also be calculated
from fluctuations
employing Callen-Welton theorem. We have
$ W^{CW}_d  = \beta\sigma^2_W/2$. The inset in Fig. (2) shows
$W^{CW}_d -  W^J_d $. This quantity is large for
small $\tau$. It decreases with increases of $\tau$
and eventually goes to zero.

\begin{figure}[htp]
\includegraphics[width=140mm]{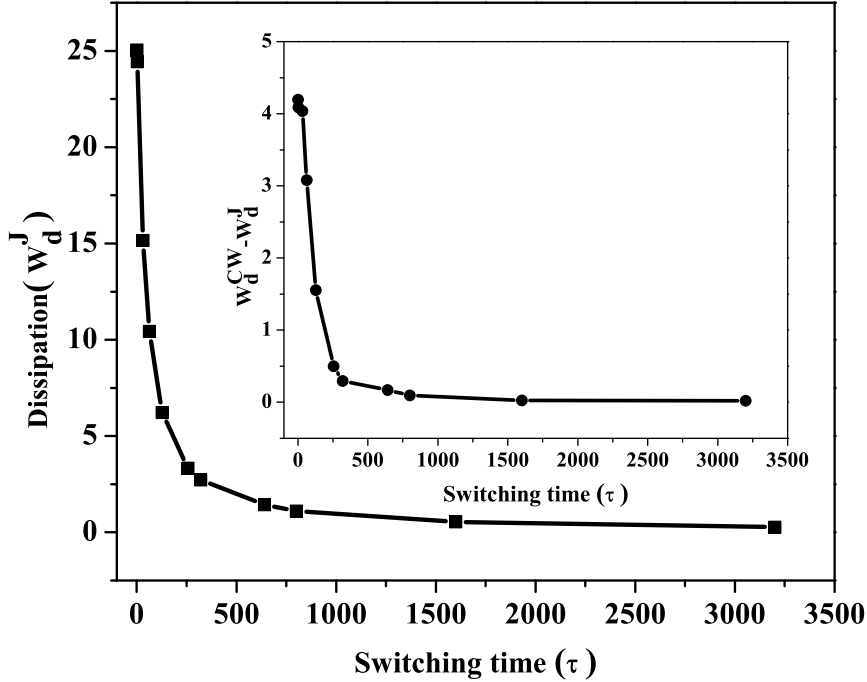}
\caption{\small{\it  Dissipation defined as
$W^J_d = \langle W\rangle - W_R$ calculated from Jarzynski identity.
The inset shows $W^{CW}_d - W^J_d$, where
$ W^{CW}_d$  is dissipation calculated from Callen-Welton theorem.}}
\end{figure}

We have depicted in Fig. (3) the probability of violation of the second law,
calculated from the Monte Carlo ensemble
of work values. Let $p^{MC}(\tau)$
denote this quantity which is
calculated as follows.

\begin{figure}[htp]
 \includegraphics[width=140mm]{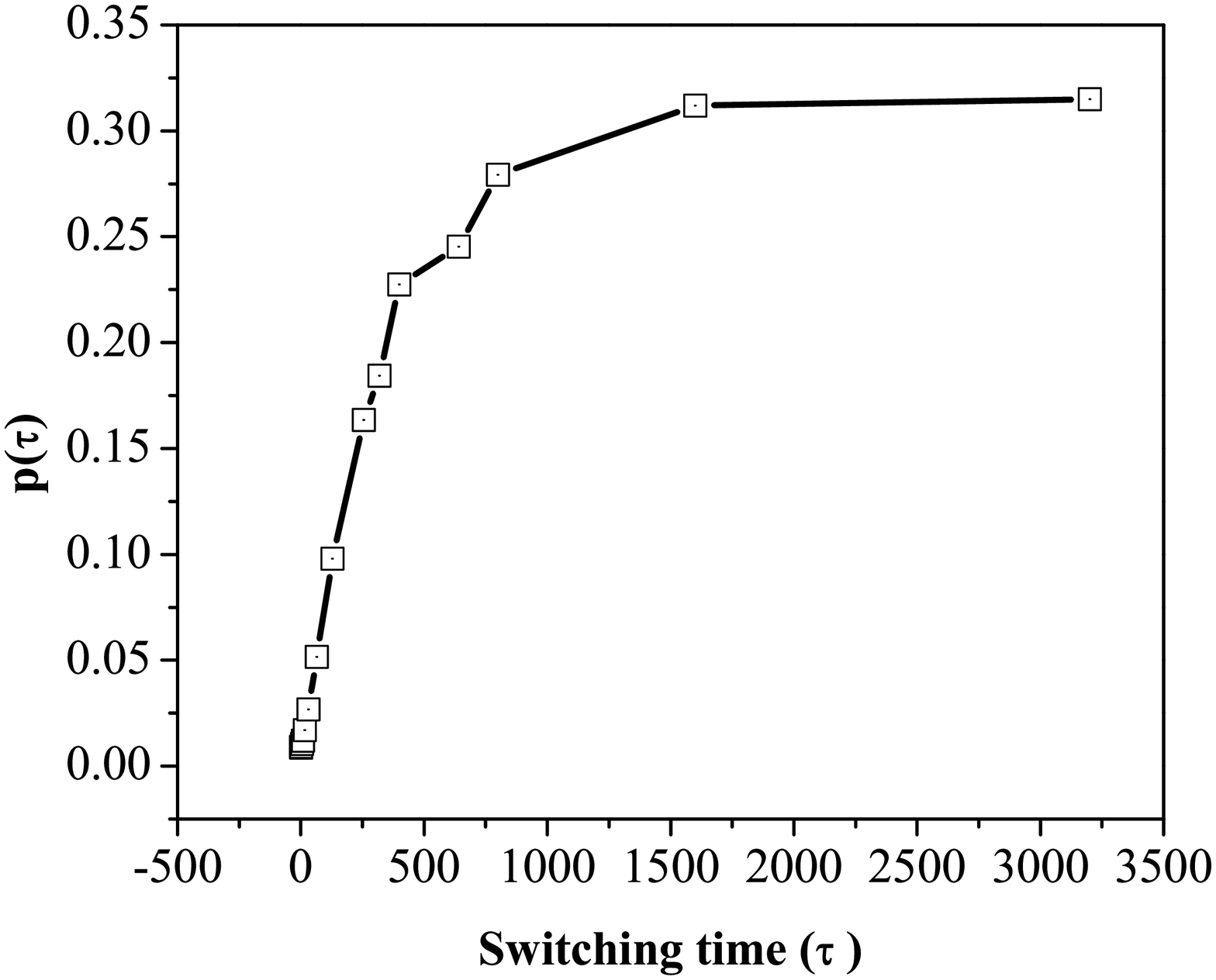}
\caption{\small{\it  The probability of second law violation for various value of $\tau$.
calculated from the Monte Carlo ensemble}}
\end{figure}

\noindent
We count how many of the switching experiments give rise to a value of
$W$ less than $W_R$. This number divided by the total number
of switching experiments carried out
gives $p^{MC}(\tau)$. We find that
$p^{MC}(\tau)$  increases with increase of
$\tau$ as expected from the analytical results discussed in section \Rmnum{5}.

The work distributions for representative values of $\tau$
are depicted in Fig. (4).
For small values of $\tau$  we find the distribution is broad.
As $\tau$ increases the distribution
becomes more and more sharply peaked. The distribution is expected to be Gaussian see \cite {Seifert,MJ}.
\begin{figure}[htp]
 \includegraphics[width=140mm]{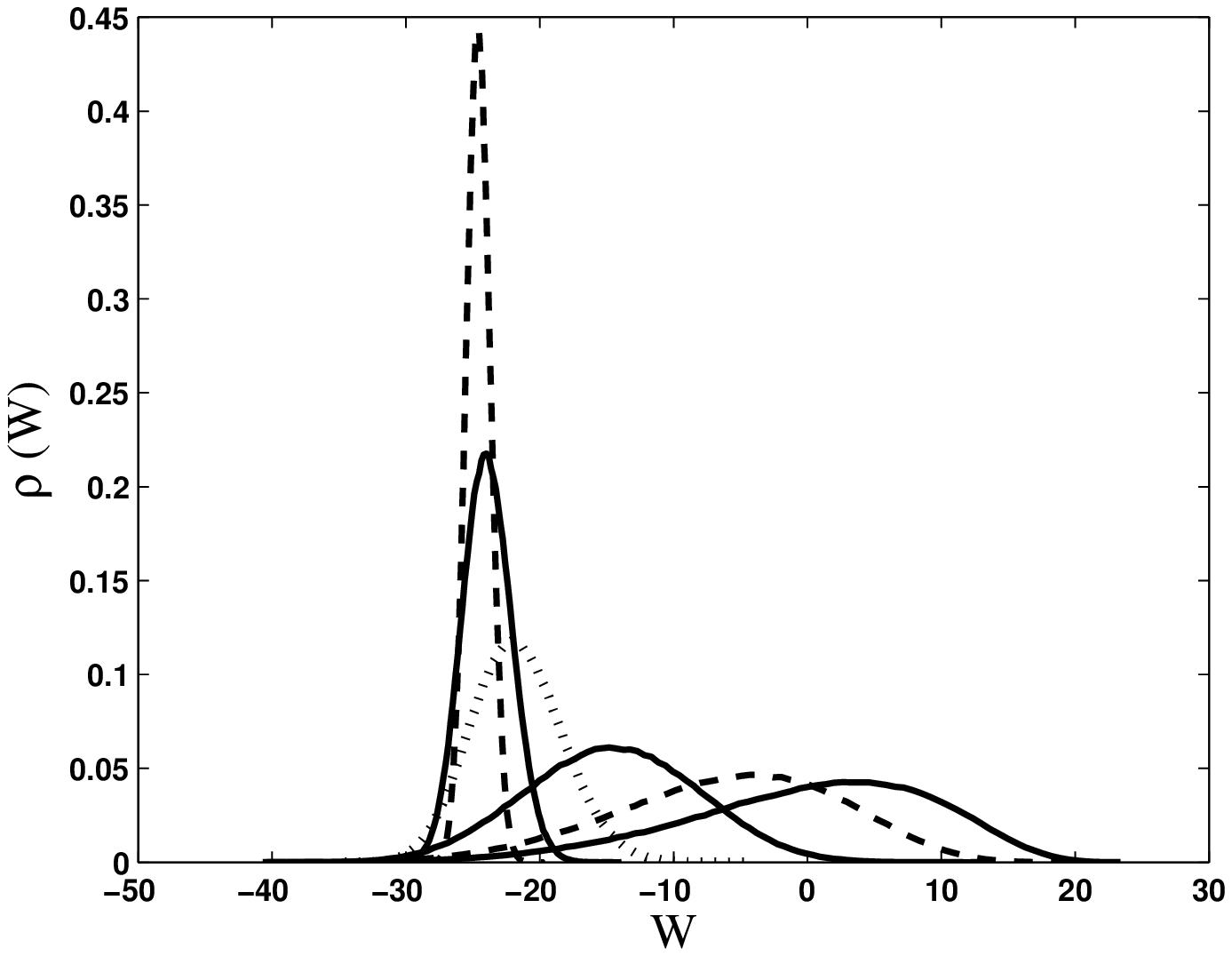}
\caption{\small{\it  Distribution of work for $\tau = 1,\ 16,\ 64,\ 256,\ 800$ and $3200$, from right to left.}}
\end{figure}


\section{Conclusions}
Recent developments in
non-equilibrium statistical mechanics
embodied in various  fluctuation theorems give us
an  insight into the
foundational aspects of statistical mechanics
and thermodynamics. There are several other methods
which provide us a
 tool to calculate equilibrium quantities  from
non-equilibrium measurements. 
For example Sadhukhan and Bhattacharjee\cite{Smb} have shown that 
Barkhausen noise process, repeated many times give  
adequate data to construct,
in conjunction with work fluctuation theorem,  
a special matrix whose principal eigenvector provides
equilibrium distribution. 
Transient entropy-fluctuation theorem
of Evans and Searles \cite{Evans}, 
heat fluctuation theorems of Crooks\cite{Crooks} 
and the fluctuations theorems of 
Gallavotti-Cohen \cite{Gallavotti}  for steady state systems  
also can be employed to 
estimate  equilibrium quantities from non-equilibrium measurements.
In this paper we have employed work fluctuation theorem
for estimating free energy differences. We have carried out
Monte Carlo simulation of the lattice model of liquid crystalline system. 
Monte Carlo is best suited for lattice spin models. If the Hamiltonian
involves continuous degrees of freedom 
then Molecular Dynamics simulations would be appropriate.
We find that dissipation defined as the excess of work
done on the system  over equilibrium free energy change
tends to zero in the asymptotic limit of $\tau\to\infty$.
$p(\tau)$ is given by
 the area under the work distribution in the tail region
 extending from $-\infty$ to $ W_R $.
We have  shown  analytically that  the value of $p$
increases with increase of $\tau$.   In the
quasistatic limit of $\tau\to\infty$, $p(\tau) =  1/2$.

\section{Acknowledgements}
MSK would like to thank Center for Advanced Studies, School of Physics, University of Hyderabad for financial support for this work.


\begin{thebibliography}{100}
\bibitem{CJ_1}
C. Jarzynski,  Phys. Rev. Lett. {\bf 78}, 2690 (1997).
\bibitem{CJ_2}
C. Jarzynski,  Phys. Rev. E {\bf 56}, 5018 (1997);
arXiv: cond-mat/9707325 .
\bibitem{J_Kurchan}
Jorge Kurchan, J. Phys. A: Math. Gen. {\bf 31}, 3719 (1998).
\bibitem{RJHGMS}
R. J. Harris and G. M. Sch\"utz, J. Stat. Mech.: Theory and Experiment,
P07020, 1742 (2007),
\bibitem{D_J_Evans}
D. J. Evans, Molecular Physics {\bf 101},  1551 (2003).
\bibitem{HBCTAW}
H. B. Callen and T. A. Welton, Phys. Rev. {\bf 83}, 34 (1951)
\bibitem{AR}
This explanation has taken from the report of an anonymous Referee.
\bibitem{JSD}
J. S. Dugdale, {\it Entropy and its Physical Meaning},
Taylor \& Francis, (1996)p.60
\bibitem{Maxwell}
James Clerk Maxwell, {\it  Tait's Thermodynamics},
Nature {\bf 17}, 257 (1878),
reprinted in ed. W. D. Niven, { \it The Scientific Papers of James Clerk Maxwell}, 
Vol. II  Cambridge: at the University Press  (1890)p.660
\bibitem{Maxwell_Demon}
J. C. Maxwell, {\it Letter to P. G. Tait, 11 Dec. 1867},
in G. C. Knot, {\it Life and Scientific Work of Peter Guthrie Tait},
Cambridge University Press, London (1924)p.213;
C. M. Caves, Phys. Rev. Lett. {\bf 64}, 2111 (1990);
C. M. Caves, W. G. Unruh and W. H. Zurek,
Phys. Rev. Lett. {\bf 65}, 1387 (1990);
 H. S.  Leff and A. F. Rex (Eds.), {\it Maxwell's Demon},
Princeton Univ. Press, Princeton (1990);
H. S. Leff, A. F. Rex (Eds.),
{\it Maxwell's Demon 2: Entropy, Classical and Quantum Information},
Computing, Institute of Physics (2003);
\bibitem{errf}
 Milton Abramowitz and Irene A. Stegun, eds.
{\it Handbook of Mathematical Functions with Formulas, Graphs, and Mathematical Tables},  New York: Dover, 1972. (See Chapter 7)
\bibitem{LL}
P. A. Lebwohl and G. Lasher, Phys. Rev. A {\bf 7}, 2222 (1973).
\bibitem{Metropolis}
N. Metropolis, A. W. Rosenbluth, M. Rosenbluth, A. H. Teller, and E.
Teller, J. Chem. Phys. {\bf 21}, 1087 (1953).
\bibitem{Barker}
J. A. Barker  and R. O. Watts, Chem.Phys.Lett., 3:144, 1969

\bibitem{Seifert}
T. Speck and U. Seifert, Phys Rev E {\bf70}, 066112, 2004
\bibitem{MJ}
O. Mazonka and C. Jarzynski, cond-mat/9912121.
\bibitem{Smb}
P.Sadhukhan and S.M. Bhattacharjee, J. Phys. A : Math. Theor. {\bf 43}, 245001 (2010)
\bibitem{Crooks}
G. E. Crooks, J. Stat. Phys. {\bf 90}, 1481 ͑(1998͒);
G. E. Crooks Phys. Rev. E {\bf 61}, 2361 (2000).
\bibitem{Evans}
D. J. Evans and D. J. Searles, Phys. Rev. E {\bf 50}, 1645 ͑(1994͒);
D. J. Evans and D. J. Searles, Phys. Rev. E {\bf 53}, 5808 ͑(1996͒).
\bibitem{Gallavotti}
G. Gallavotti and E. G. D. Cohen, Phys. Rev. Lett. {\bf 74}, 2694 (͑1995͒);
G. Gallavotti and E. G. D. Cohen, J. Stat. Phys. {\bf 80}, 931 (͑1995͒).
\end{thebibliography}
\end{document}